\documentclass[1p,12pt]{elsarticle}

\journal{Nuclear Instruments And Methods A}

\usepackage[pdftex]{color}                  
\usepackage{graphics}               
\usepackage{fullpage}
\usepackage{setspace}
\usepackage{verbatim}
\usepackage{hyperref}

\begin{document}
\begin{frontmatter}
\title{Characterization of the Hamamatsu R11780
	12 inch Photomultiplier Tube}

\author[csu]{J. Brack}
\author[penn]{B. Delgado}
\author[davis]{ J. Dhooghe}
\author[davis]{J. Felde}
\author[csu]{B. Gookin}
\author[penn]{S. Grullon\corref{cor1}}
\ead{grullons@hep.upenn.edu}
\author[penn]{J.R. Klein}
\author[nor]{R. Knapik}
\author[chi]{A. LaTorre}
\author[penn]{S. Seibert}
\author[penn]{K. Shapiro}
\author[davis]{R. Svoboda}
\author[penn]{L. Ware}
\author[penn]{R. Van Berg}

\address[penn]{University of Pennsylvania, Philadelphia, PA 19104, USA}
\address[nor]{Norwich University, Northfeild, VT 05663, USA}
\address[chi]{University of Chicago, Chicago, IL 60637, USA}
\address[csu]{Colorado State University, Fort Collins, CO 80523, USA}
\address[davis]{University of California, Davis, Davis, CA 95616, USA}

\cortext[cor1]{Corresponding author}


\begin{abstract}

	Future large water Cherenkov and scintillator detectors have been proposed for
measurements of long baseline neutrino oscillations, proton decay, supernova
and solar neutrinos.  To ensure cost-effectiveness and optimize scientific
reach, one of the critical requirements for such detectors are large-area, high
performance photomultiplier tubes (PMTs).  One candidate for such a device is
the Hamamatsu R11780, a 12'' PMT that is available in both standard and high
quantum efficiency versions.  Measurements of the single photoelectron response
characteristics, relative efficiencies of the standard and high quantum
efficiency versions, a preliminary measurement of the absolute quantum efficiency of the standard quantum efficiency version,
 and a two-dimensional scan of the relative efficiency across the photocathode surface are presented in this
paper.  All single photoelectron investigations were made using a Cherenkov light source at room
temperature at a gain of $1\times10^{7}$.  These results show that the R11780 PMT is a excellent candidate
for such large optical detectors.

\end{abstract}


\end{frontmatter}
\section{Introduction}

	Great discoveries in neutrino physics have been made with large,
monolithic, optical detectors.  The Super-Kamiokande (Super-K) 40~ktonne water
Cherenkov detector discovered unambiguous evidence of atmospheric neutrino
oscillations~\cite{superk}, the Sudbury Neutrino Observatory (SNO) solved the
long-standing solar neutrino problem using a 1 ktonne heavy-water Cherenkov
detector~\cite{snoprl1} and the KamLAND $\sim$1~ktonne liquid scintillator
detector observed the same oscillation terrestrially with reactor
antineutrinos. Recently the RENO~\cite{reno} and Daya Bay~\cite{dayabay} liquid
scintillator detectors have made definitive measurements of  the third
neutrino mixing angle. In each of these experiments, the performance of large-area
photomultiplier tubes (PMTs) was critical to ensure large scientific reach at
reasonable cost.

	It is likely that further progress in neutrino physics can be made
using large water or liquid scintillator detectors, and several proposals for
such detectors have been put forward~\cite{lbnecdr,lena,hyperk,deepcore}.   In
addition, searches for proton decay either via $p\rightarrow e^+\pi^0$ or,
possibly with scintillator or loading with Gd~\cite{gadzooks}, $p\rightarrow
K^+\bar{\nu}$, as well as studies of atmospheric neutrinos including searches
for non-standard interactions, solar neutrinos including observation of the
Day/Night asymmetry~\cite{baltz}, and possibly even, in loaded scintillator,
searches for neutrinoless double beta decay (following the SNO+~\cite{snot} and
KamLAND-Zen~\cite{kamzen}) can be done with very large masses observed by
efficient, inexpensive PMTs.  The IceCube Neutrino Observatory \cite{icecubeperformancepaper} consists of a kilometer-cubed instrumented volume of glacial ice at the South Pole with well-tested large area 10-inch PMTs \cite{icecubepmtpaper}. Very large-scale `single-phase' dark matter
detectors~\cite{clean} have also been proposed that use the same principle.

	Typically, the physics reach of such detectors is optimized by
increasing the number of detected photons, narrowing the resolution for photon
arrival times, and being able to distinguish single photons from multiple
photons.  These general requirements argue for a very large number of PMTs with
excellent timing and charge response. Maintaining a reasonable cost, however,
argues for reducing the total number of channels, by using the largest-area
phototubes that can withstand the environment (e.g., pressure and temperature
within the detector volume), and by increasing the efficiency of each PMT.
	
The 12-inch Hamamatsu R11780 is an excellent candidate for such large detectors
based on its large size, the existence of a high quantum efficiency (HQE)
design, and a mechanical construction that can withstand high water
pressures \cite{lbnewatercdr}.  We describe here measurements characterizing the single
photoelectron response of both the standard and high quantum efficiency
versions of the Hamamatsu R11780 PMT.  Section \ref{pmtchap} describes the Hamamatsu R11780 PMT and describes magnetic field effects on the PMT response. Section \ref{spechap} describes the experimental setup used for the single photoelectron characterization.  We describe the results of the charge and time characterization of the R11780 PMT in Section \ref{sec:charge} and Section \ref{sec:TTS}, respectively, a measurement made of the absolute quantum efficiency in Section \ref{sec:eff}, and describe the position dependence along the photocathode of the detection efficiency and timing response in Section \ref{sec:2dscans}.

\section{The R11780 PMT \label{pmtchap}}

\begin{figure}[hbt!]
   \centering
    \includegraphics[width=0.6\textwidth]{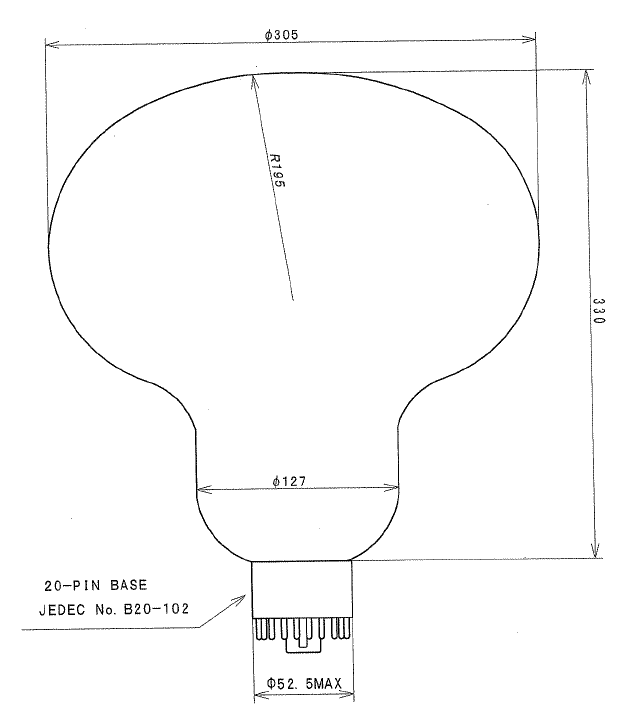}\\
\includegraphics[width=0.4\textwidth, angle=90]{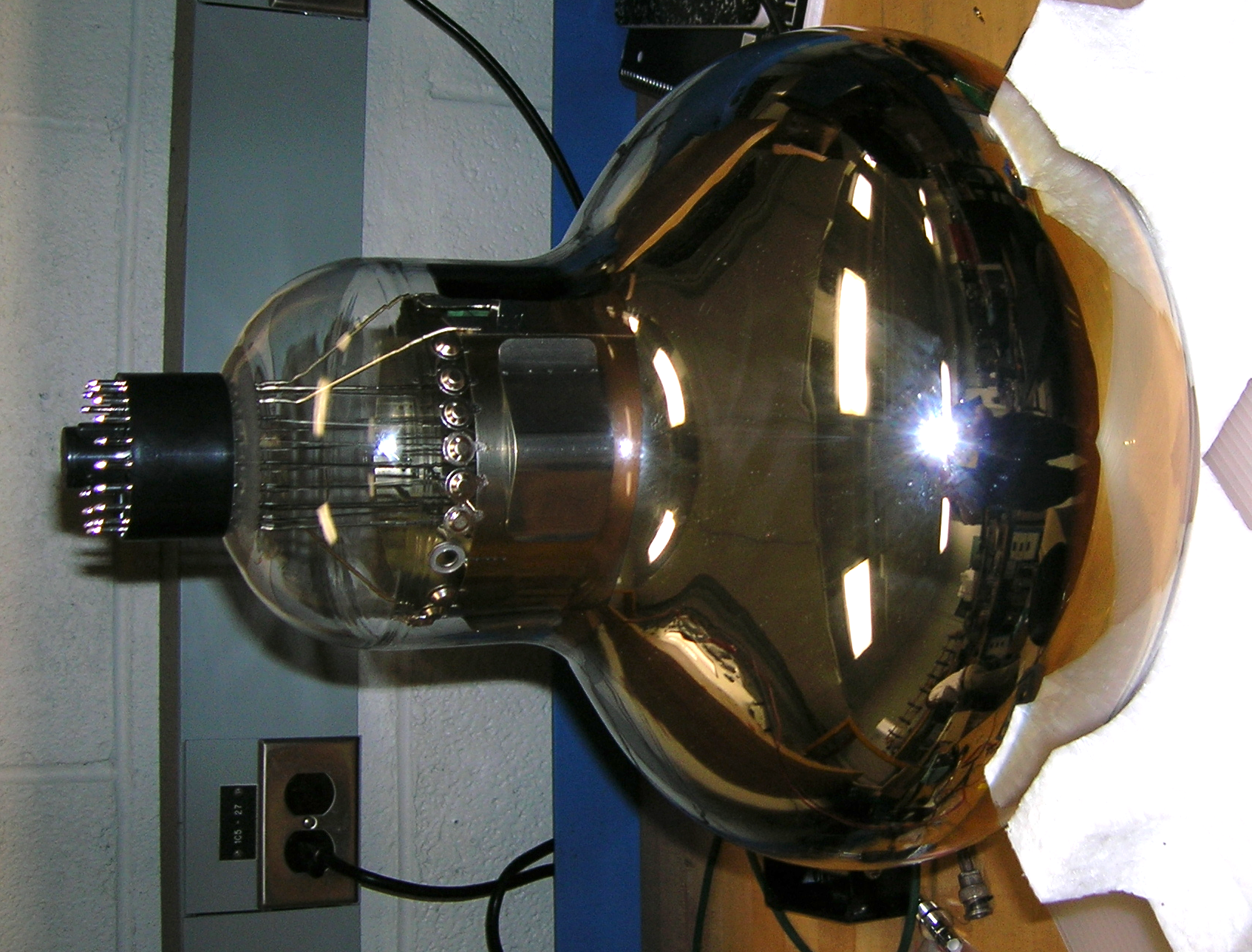}	
\includegraphics[width=0.2\textwidth]{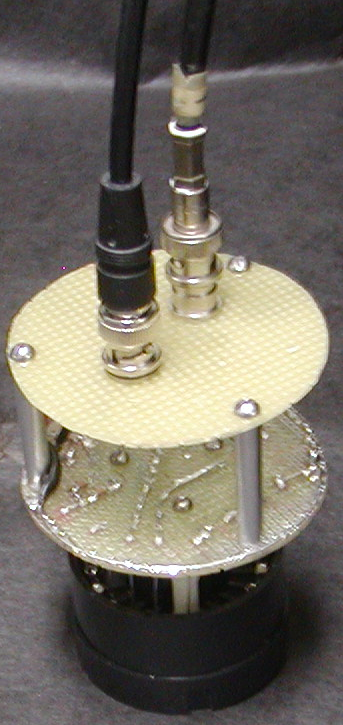}
    \caption{ The R11780 12 inch Hamamatsu PMT and the two-cable voltage divider used. The two-cable voltage divider was designed to be cathode-grounded, and the  coupling capacitor required to separate the signal from the positive high voltage was included for convenience. }
    \label{R11780}
\end{figure}


The R11780 PMT developed by Hamamatsu Photonics (Figure \ref{R11780}) has a 12''
diameter photocathode, and ten linear-focused dynode stages.  We tested two
sets of phototubes: ``standard quantum efficiency" (SQE) tubes with quantum
efficiencies anticipated to peak near 21\% at 390~nm,  and bialkali ``high
quantum efficiency" (HQE) tubes with quantum efficiencies peaking around 32\%
at 390~nm.  We note that the anticipated peak quantum efficiency of the standard configuration is lower than the average peak quantum efficiency for the Hamamatsu 10-inch R7081 PMT, which is typically $25\%$ \cite{icecubepmtpaper}, and that the anticipated peak quantum efficiency of the high quantum efficiency configuration is the same as the average peak quantum efficiency for the high quantum efficiency configuration of the Hamamatsu R7081 PMT of $32\%$ at 390 ~nm \cite{deepcore}.  


\subsection{Measurement of Effects due to Magnetic Fields}
It is well known that the electron focusing capabilities of a PMT are affected by
magnetic fields. The Lorentz force felt by the electron implies that a field
transverse to the PMT will tend to deflect
some fraction of electrons away from the first dynode, thereby reducing the PMT's
collection efficiency \cite{hamamatsuhan}.  Tests to investigate the magnitude of this effect were done for
one R11780 high quantum efficiency PMT in a tunable, tri-axial, magnetic field environment at the University
of California, Davis. Three pairs of 1~m$^{2}$ copper Helmholtz coils, separated
by 0.5~m, were arranged along perpendicular axes inside a dark room. Each coil pair was
connected in series, and a variable DC current, capable of producing fields of $\sim$500~mG
(similar to the magnitude of the Earth's magnetic field), was driven through each. 
The uniformity of the field was measured to
be $\pm$ 10~mG within a 12" cube around the PMT bulb. This non-uniformity is primarily
caused by variations in the Earth's magnetic field inside the laboratory.

During testing it was discovered that the R11780's purchased were constructed
with a metal tab of considerable magnetization affixed to the outer base of the dynode
stack. This magnetization is localized, but does introduce some field non-uniformity within
the PMT. The fields are measured at the location of the center of the PMT prior
to placing the PMT inside the coils to accurately simulate external field values.

Testing was carried out by initially canceling the vertical component of the Earth's magnetic
field, and then varying independently the field strength in both transverse directions. For
these studies, the PMT was supported such that it was pointed upward, in the +Z direction.
The +X direction is defined by the notch on the PMT's base connection, between the
first and twentieth pins, and the +Y direction follows a right handed coordinate system
convention. The light source used was a green (570~nm) LED positioned $\sim$0.5~m above
the PMT to fully illuminate the PMT face. Since the magnetic field affects the electron 
focusing, dependence on the photon wavelength is assumed to be small, and ultimately not easily mitigated by external means. 

Data was acquired using a Tektronix DPO7254 digitizing oscilloscope. At each
field value the individual PMT waveforms from 50,000 identical LED triggers are stored .
The waveforms are integrated in software to remove small DC offsets, and obtain the
average charge collected. During these tests the PMT was operated at a gain of $1\times10^{7}$.

\begin{figure}[htbp]
   \centering
   \includegraphics[scale = 0.57]{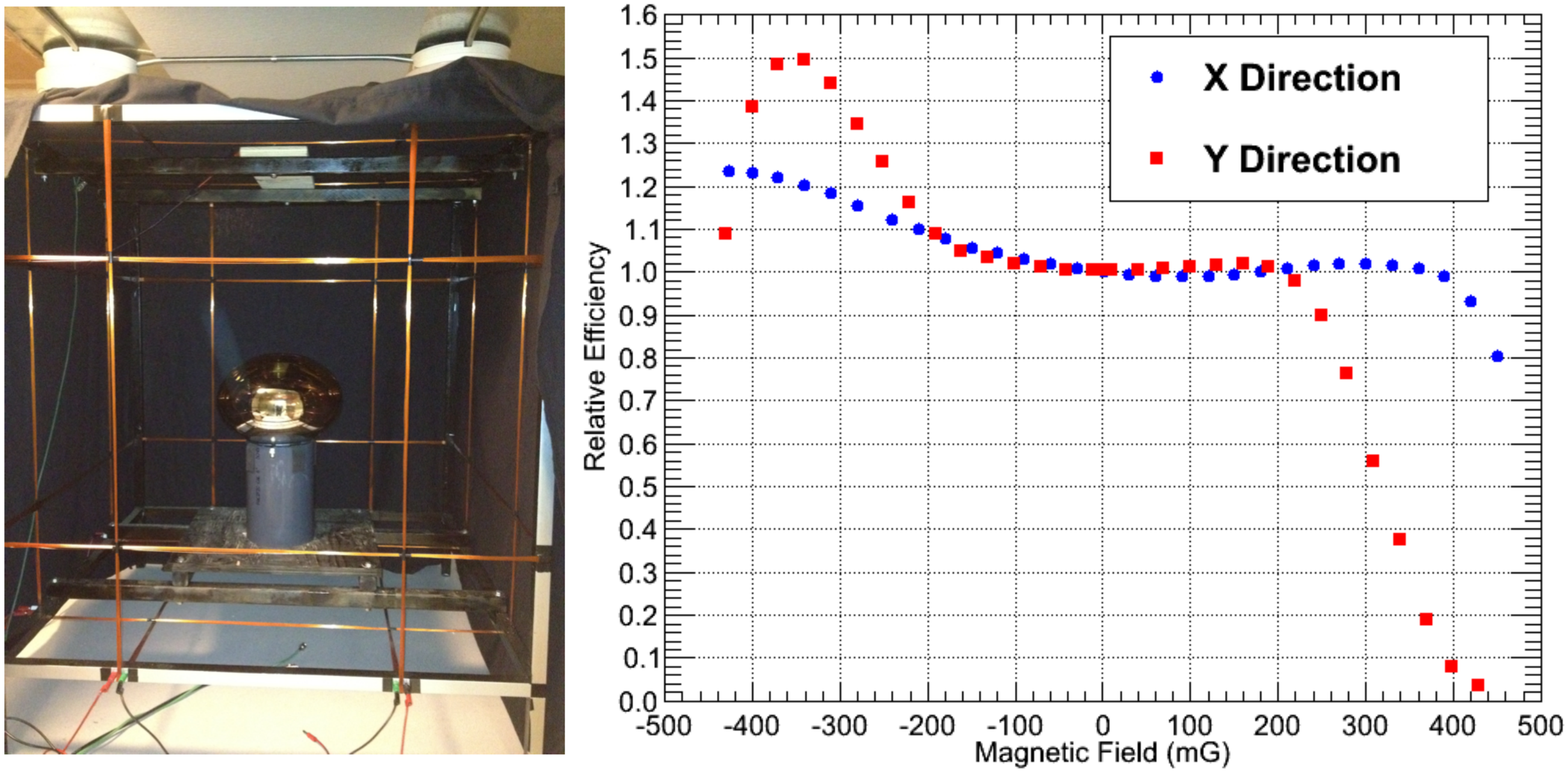}
   \caption{Left: Helmholtz coils and PMT testing setup at UC Davis. Right: Results of
   measured charge relative to zero field for a number of magnetic field strengths.
   During each test the opposite transverse field value is held at zero.}
   \label{fig:example}
\end{figure}

A fine scan was done for magnetic fields along X and Y between -450~mG and +450~mG.
The data show an expected degradation of efficiency at fields along the +X and +Y
directions, but an unexpected increase in efficiency for fields in the -X and -Y directions. One
can see, however, that for fields greater than 500~mG the data appear to return to the
expected behavior of decreasing efficiency.  Although this behavior is unexpected, it can be understood that the PMT collection efficiency is not optimized with respect to this effect.It should be noted that these measurements can not be used to accurately predict the effect of a field along X and Y simultaneously, but the
scale of the effect is well established by the data presented.

In a future large scale detector instrumented with R11780 PMTs, it is clear that mitigation of
the Earth's magnetic field is necessary.  In some cases, simply orienting the PMT in an
optimal way could be sufficient. To maximize PMT performance, it is usually desirable to
either actively cancel the field with coils, or passively reduce the field using a high magnetic
permeability metal surrounding parts of the PMT. These tests indicate that transverse fields
$<$ $\sim$200 mG will have $<$ $\sim$5\% degradation, and can even result in improved
performance.

\section{Single Photoelectron Characterization Experimental Setup \label{spechap}}

 For a large-scale Cherenkov or scintillator experiment, the detector's
capability for discriminating different particle types, reconstructing particle
positions and directions, and measuring energy deposition depends primarily on
the performance of the chosen PMT.  The PMTs in such large detectors typically
detect no more than a few photons per event.  Thus, the single photoelectron response
is most important in determining the optimal PMT. A detailed characterization
of this response is also critical for developing a complete model of the
detector, useful both for understanding the detector's response and for
position, energy and direction reconstruction in the case of Cherenkov detectors.

\begin{figure}[hbt!]
   \centering
    \includegraphics[width=0.5\textwidth]{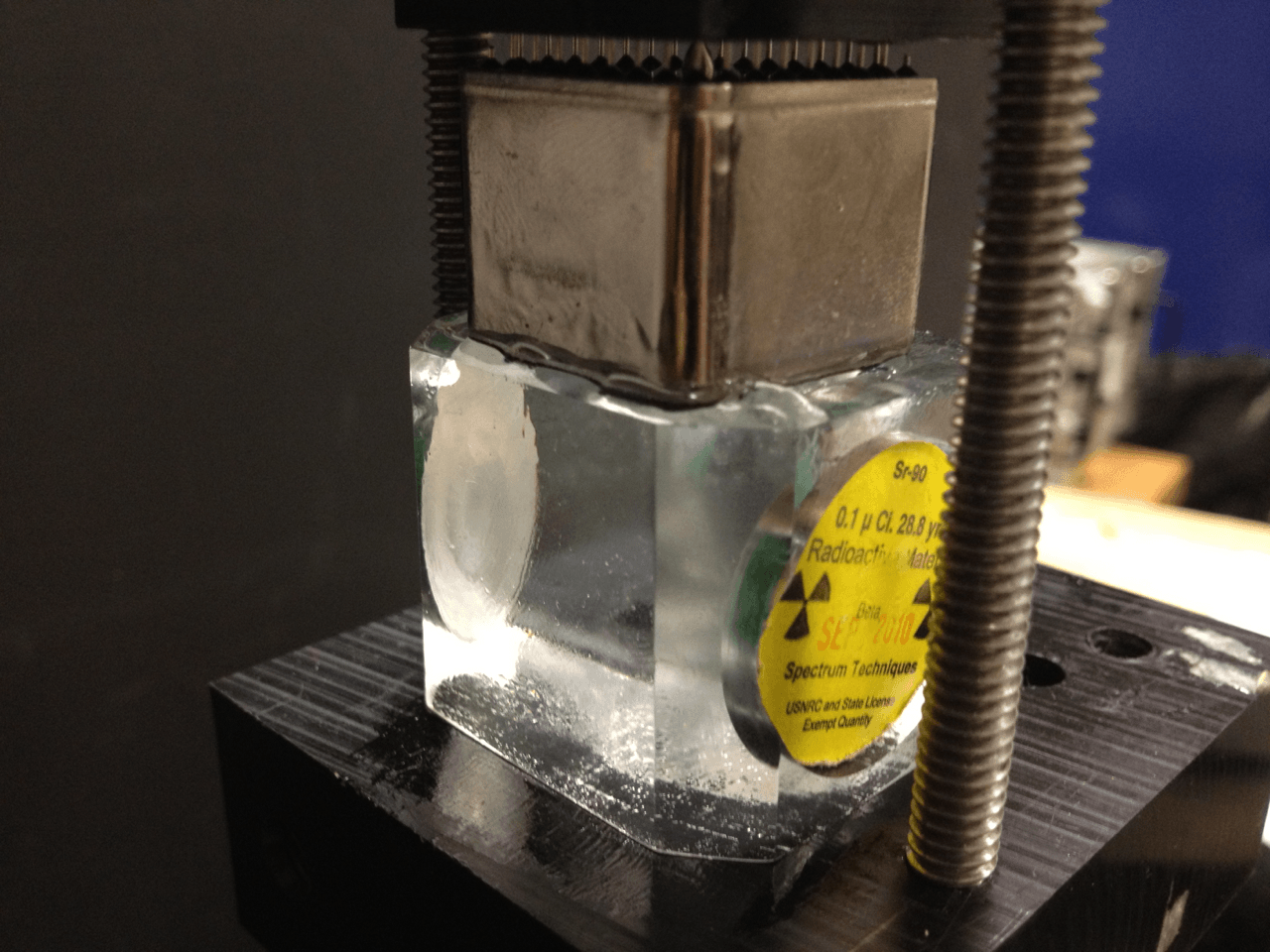}
    \caption{ The Cherenkov light source. Two 0.1 $\mu$Ci $^{90}$Sr discs are embedded on the side of a 1 inch cube, which is optically coupled to a 1 inch PMT on top of the acrylic that acts as a trigger.}
    \label{TriggerCube}
\end{figure}

   Our measurements of the single photoelectron response of the Hamamatsu
R11780 uses a simple Cherenkov light source.  The Cherenkov source was constructed by
embedding two 0.1 $\mu$Ci discs of $^{90}$Sr in a piece of UV-transparent
acrylic machined into an approximately 1 inch cube.  The acrylic is the same
material as was used in the acrylic vessel (the acrylic sheets were provided by PolyCast\textsuperscript{\textregistered})  to hold the heavy water in the
Sudbury Neutrino Observatory (SNO) experiment, and its optical properties have
been already studied in great detail \cite{sno}.  A 1-inch Hamamatsu
R7600-200 super bialkalai high quantum efficiency PMT is optically coupled with
the acrylic source and is used as a fast ($\sim$250~ps FWHM) trigger as shown in Figure
\ref{TriggerCube}.
The $^{90}$Sr undergoes a 0.546 MeV $\beta$- decay into the yttrium isotope
$^{90}$Y with a half-life of 29.1 years. The $^{90}$Y also undergoes a $\beta$-
decay into $^{90}$Zr (zirconium), which is stable.  The half life for $^{90}$Y is 64 hours
and the emitted electron from the $^{90}$Y  decay has an energy 2.28 MeV which will enter the acrylic and will produce Cherenkov light.  

	The Cherenkov source has the advantage that its wavelength spectrum is
broadband, nearly identical to a water Cherenkov detector and spanning the
same range as a scintillation detector.  The photons are produced with a very narrow time distribution, and thus allow
precision measurements of the PMT timing response.  With the PMTs placed 45~cm
from the Cherenkov source, the fraction of coincidences between an individual
PMT and the trigger PMT was $\sim$3-5\%. Assuming the number distribution of photons incident on the PMT photocathode are Poisson distributed, we were predominantly in the single photoelectron regime. Nevertheless, there was a small contamination from multiple photons that are not described by Poisson statistics. This non-Poisson contribution of multiple photons is due to the correlated nature of Cherenkov light, cases where the Cherenkov cone points primarily toward the PMT and therefore is not a pure source of single photoelectrons. We have not made any corrections for that effect in the measurements we present here because the SPE characterization is insensitive to this effect.  Section
\ref{sec:sys} shows this contamination is relatively small and only
affects the high tail of the charge distribution.

\begin{figure}[hbt!]
   \centering
    \includegraphics[width=0.9\textwidth]{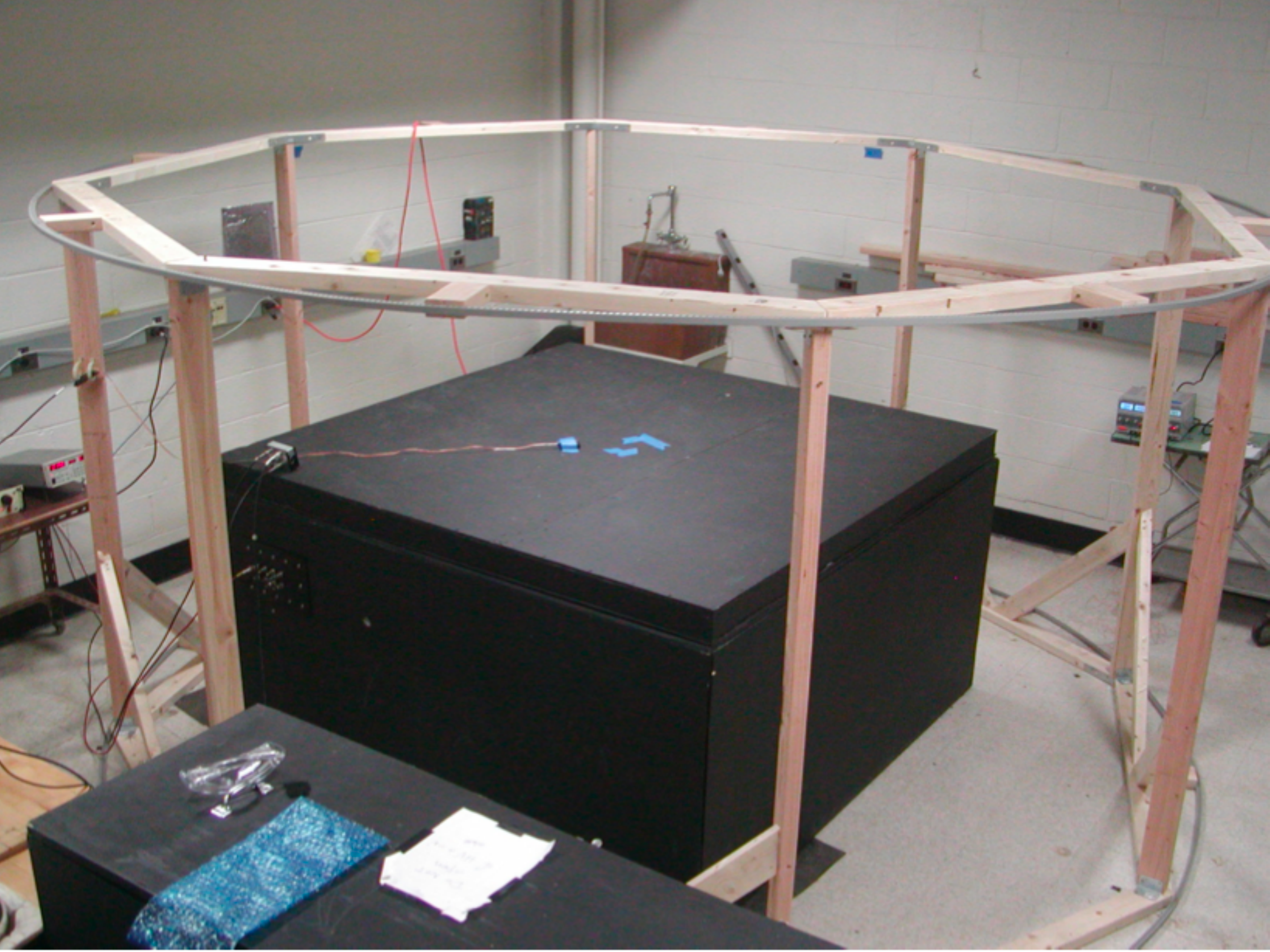}
    \caption{ The dark box setup with Helmholtz coils for cancellation of the Earth's magnetic field.  At  a distance of 1.5~m,
	      the coincidence rate is kept below 5\%.}
    \label{dark_box_wide}
\end{figure}

A 2~m by 2~m by 0.9~m dark box (Figure~\ref{dark_box_wide}) was built to house
the Cherenkov light source along with the PMTs being tested. Our data
acquisition system was a LeCroy Waverunner 6050A 500~MHz  oscilloscope, running
LeCrunch~\cite{lecrunch} software that allowed us to acquire full waveform data  
at rates up to 1~kHz.  No additional amplification of the PMT signal was performed. Given the size of the dark box and the speed of our
acquisition system, we were able to characterize up to 3 PMTs at a time, each
optically isolated from one another.  The entire dark box
setup is surrounded by a set of  Helmholtz coils to help reduce the effects of
the vertical component of Earth's magnetic field.  After compensation, the
magnetic field in the dark box has a upward z-component of less than 5~mG and
residual components in the x and y directions of around 170~mG and 50~mG
respectively.  For some of our measurements, we also added Finemet\textsuperscript{\textregistered} \cite{finemet}
shielding to the entire box, further reducing the $x$ and $y$ components of the
field to 34 ~mG and 23 ~mG, respectively.

\subsection{Data Acquisition and Storage \label{lecroy}}

\begin{figure}[hbt!]
   \centering
    \includegraphics[width=0.9\textwidth]{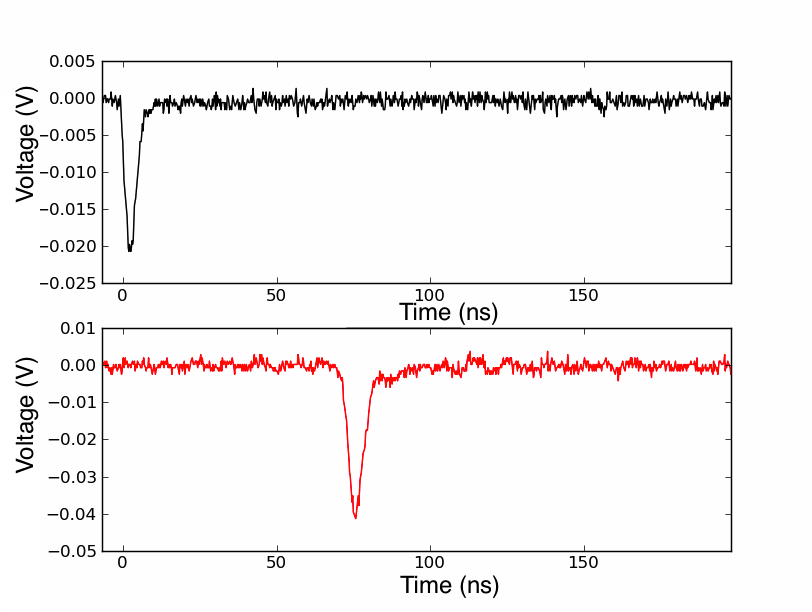}
    \caption{ The entire trigger PMT digitized trace is displayed (top), as well
	     as the trace for the R11780 PMT being characterized (bottom).  The waveforms 
	     were used for both timing and charge.  Due to the low coincidence rate, the majority of triggers recorded no pulse from the R11780 PMT.}
    \label{trace_one}
\end{figure}

Full waveform traces were stored for both the trigger PMT and the PMTs being
tested.  Each trace consists of 1252 samples at 0.2~ns intervals for a total of 250.4~ns per trace. Figure~\ref{trace_one} shows typical waveform traces in a
coincident event for a R11780-SQE PMT.  

The raw data were transferred to a data storage server in a custom HDF5
format. For a gain of $1\times10^{7}$, 20 individual data runs were recorded to provide adequate statistics in the single photoelectron
(SPE) analysis.  A typical individual data run consisted of 100,000 triggered events which
corresponded to $\sim$3000-5000 coincident events per run.  Analysis software
processed the data immediately after the data runs were completed and the
results were uploaded into a CouchDB database.   With approximately 60,000 coincident pulses, the statistical uncertainties
     on most SPE parameters are at or below the 1\% level and therefore were sufficient for most of
the characterization.  A CouchApp web-based application was created to make
data publicly accessible \cite{couchwb} and the results were aggregated into easily accessible
tables.  


\section{R11780 Single Photoelectron Characterization Results}

\subsection{R11780 Charge Response}
\label{sec:charge}

\begin{figure}[hbt!]
   \centering
    \includegraphics[width=0.48\textwidth]{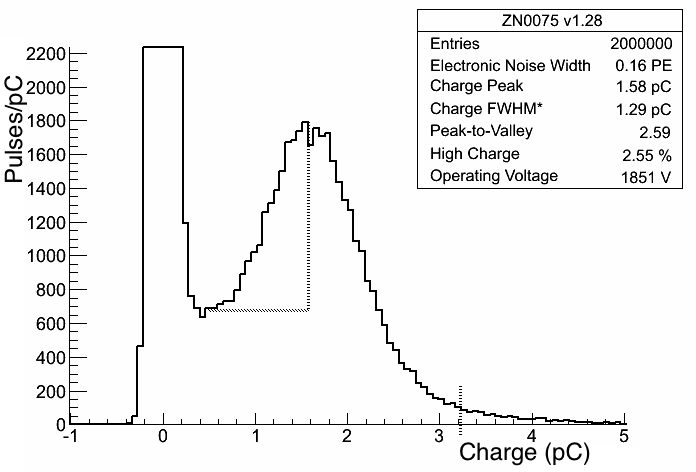}
        \includegraphics[width=0.48\textwidth]{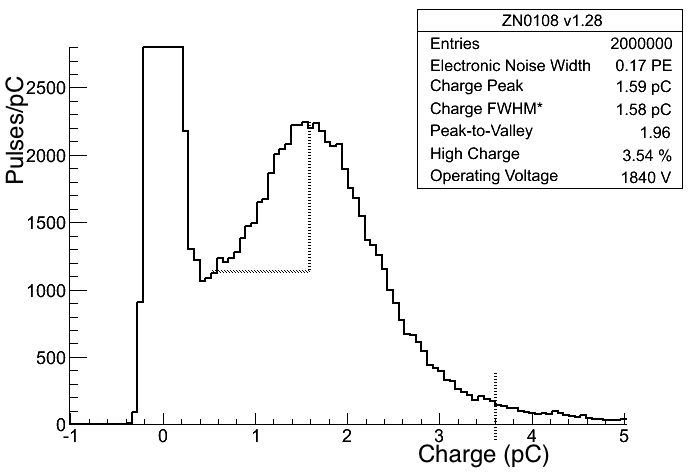}%
    \caption{ The single photoelectron spectrum for a Hamamatsu R11780 PMT.  The
	      left distribution is displayed for a standard quantum efficiency PMT and on the right  
	      is shown for the high quantum efficiency model. The events to the right of the dashed vertical 
	      line at $\sim$3~pC represents the ``high charge tail" of the distribution. The enhancement near 3.2 pC may be due to events with two photoelectrons.  The high charge tails in the two distributions are higher than expected from Poisson statistics, for the reasons outlined in the text. The expected number of multiple photoelectron events from Poisson statistics for these examples is negligible since the coincidence rate for the left distribution is $1.2\%$, and the coincidence rate for the right distribution is $2.0\%$ (which correspond to Poisson means of 0.012 and 0.02, respectively).
  }
    \label{fig:charge_png}
\end{figure}

	Figure~\ref{fig:charge_png} shows typical SPE spectra for both the
standard and high quantum efficiency R11780 PMT models. The data were taken using the
triggered Cherenkov source.  These spectra were created by integrating over a
fixed 30~ns `prompt' time window in each R11780 waveform for all triggered
events.  

The figures that demonstrate the SPE charge
response of a PMT in this paper, such as Fig.~\ref{fig:charge_png},  are broadly characterized by the
convolution of two distributions.  The first is a tall narrow distribution
centered at 0.0~pC, which is the electronic noise inherent in the data
acquisition system.  The second is the actual SPE charge distribution with
a peak at about 1.6~pC, which corresponds to a gain of $1\times10^{7}$.  The height of the SPE peak relative to the depth
of the `valley' between the SPE charge distribution and the electronic
noise peak  known as the `pedestal'  provides a metric for the PMT's ability to
distinguish true photons from electronic noise.  Real photons can occasionally produce very
low charges and therefore the valley consists of
both very low charges from photoelectrons and upward fluctuations of the
electronic noise.  A larger Peak-to-Valley ratio results in a lower percentage of
low charge events and a higher overall detection efficiency for a given
charge threshold. Some PMTs exhibit a unusually long tail near the high side of the charge spectrum.
 Some of the events in the high charge tail are from multiple photoelectrons and this is discussed in
Section~\ref{sec:sys}.  The size of the high-charge tail coupled with the
width of the charge distribution affects the PMT's ability to distinguish
between single and two or more photons.
	
	For each run, we perform several fits to the charge spectra to extract
the SPE parameters.  A Gaussian is fit to the electronic noise peak, another
Gaussian is fit to the SPE distribution and a third fit is used to locate the
valley.  While the SPE distribution generally has significant non-Gaussian tails, the Gaussian approximation is adequate near the peak of the distribution for the application of comparing the SPE response of different PMTs. 
The explicit definitions of the fit parameters are:

\begin{itemize}
    \item[] \textbf{Electronic Noise Width:} The Gaussian fit is done around 
    the peak value (located near 0.0~pC) $\pm$~$N_{half}$ bins, where
    $N_{half}$ is the number of bins to the half height location on the
    positive side of the peak.  The width is defined as the sigma from the fit (4 times this value is generally
    close to 1/4~PE discriminator threshold typically used in experiments). 

    \item[] \textbf{Charge Peak:} The Gaussian fit is centered on the SPE
distribution peak, from $\frac{2}{3}$Max$_{SPE}$ to 1.5Max$_{SPE}$, where
Max$_{SPE}$ is the bin of maximum value in the integrated charge peak.  The
gain is defined as the ratio of the electron charge to the peak charge, thus a
gain of 1.0$\times10^7$ corresponds to a charge peak of 1.6 ~pC if no additional amplification is performed. Our test setup does not include an additional amplifier. 

    \item[] \textbf{Charge FWHM:} $2\sigma_{SPE}\sqrt{2ln(2)}$ is the charge
full width at half maximum value, where $\sigma_{SPE}$ is the sigma from the
Gaussian fit to the SPE distribution.  
    
    \item[] \textbf{Peak/Valley:} The height of the peak is defined by the
overall normalization of the Gaussian fit to the SPE distribution.  The height
of the valley is found by fitting a quadratic from 6$\sigma_{elec}$ to
$\frac{2}{3}$Max$_{SPE}$, and the ratio is defined between the two.  

    \item[] \textbf{High Charge Tail:} We estimate any anomalously high charge
tail by measuring the ratio of the number of events with a charge
3$\sigma_{SPE}$ above the SPE peak to the total number above the electronic
noise width (defined above).  For our Cherenkov source, the high charge tail is
non-trivially contaminated by multi-photoelectron events, and we do not correct
for that effect here.  
\end{itemize}

	The results of our measurements of the single photoelectron charge
response for standard quantum efficiency versions of the R11780 are shown in
Table~\ref{tbl:qresultsstd}, and for the high quantum efficiency versions in
Table~\ref{tbl:qresultshqe}.  The results for the standard QE tubes are averaged over seven PMTs and the HQE results are averaged over ten PMTs. The data
was taken at a gain of $1\times 10^7$ for both configurations.  As the tables show, the charge response of
these tubes is very good overall compared to the 8-inch tubes used for SNO \cite{sno} and the 20-inch tubes used for Super Kamiokande \cite{superk} .  We note again that the high charge tail has
a non-trivial contribution from multi-photoelectrons due to the nature of our
Cherenkov source and that the effect is bigger for the HQE PMTs due to
their higher detection efficiency. We also note that the Peak/Valley ratio for
the HQE PMTs appears to be systematically lower than for the standard quantum
efficiency versions, although this difference is not inconsistent with
PMT-to-PMT variations seen within each sample. 

\begin{table}[htb!]
\renewcommand{\arraystretch}{1.0}
\centering
  \begin{tabular}{c c c c c}
 \hline
  & Average & Standard Deviation &  Minimum & Maximum\\
 \hline
Charge FWHM (pC) & 1.42 & 0.4 & 1.18 & 2.32 \\
Peak/Valley &  2.8 & 0.28 & 2.3 & 3.0\\
High Charge tail (\%) & $2.86\%$ & $0.84\%$ & $2.5\%$ & $4.94\%$ \\
Poisson Mean & 0.015 & 0.006 & $0.011$ & $0.025$ \\
Operating Voltage (V) & 1848 & 75 &  1740 & 1920  \\
 \hline
\end{tabular}
\caption{Summary of PMT SPE charge characteristics results for the R11780 standard quantum efficiency PMT. The operating voltage was adjusted to achieve a gain of $1\times10^{7}$, which corresponds to a SPE charge peak of 1.6~pC. All measurements were made at room temperature.}
\label{tbl:qresultsstd}
\end{table}

\begin{table}[htb!]
\renewcommand{\arraystretch}{1.0}
\centering
 \begin{tabular}{c c c c c}
 \hline
  & Average & Standard Deviation &  Minimum & Maximum\\
 \hline
Charge FWHM (pC) & 1.64 & 0.62 & 1.19 & 3.36\\
Peak/Valley &  2.24 & 0.27 & 1.78 & 2.76\\
High Charge tail (\%) & $3.75\%$ & $0.66\%$ & $2.73\%$ & $5.2\%$ \\
Poisson Mean & 0.023 & 0.005 & $0.015$ & $0.029$ \\
Operating Voltage (V) & 1950 & 221 & 1750 & 2500 \\
 \hline
\end{tabular}
\caption{Summary of PMT SPE charge characteristics results for the R11780 high quantum 
efficiency PMT. The operating voltage was adjusted to achieve a gain of
$1\times 10^{7}$, which corresponds to a SPE charge peak of 1.6~pC. All measurements were made at room temperature.}
\label{tbl:qresultshqe}
\end{table}

\subsection{R11780 Timing Response}
\label{sec:TTS}

The transit time between the absorption of a photon at the photocathode and the
output pulse from the anode of a PMT varies from photoelectron to
photoelectron, and the inherent fluctuation in the transit time can greatly
affect position reconstruction algorithms.  A candidate PMT should therefore
have a transit time spread that is as narrow as possible.

\begin{figure}[hbt!]
   \centering
    \includegraphics[width=0.5\textwidth]{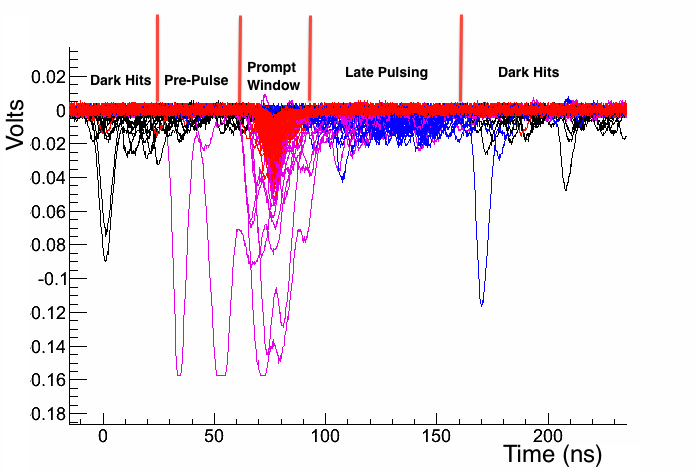}%
    \includegraphics[width=0.5\textwidth]{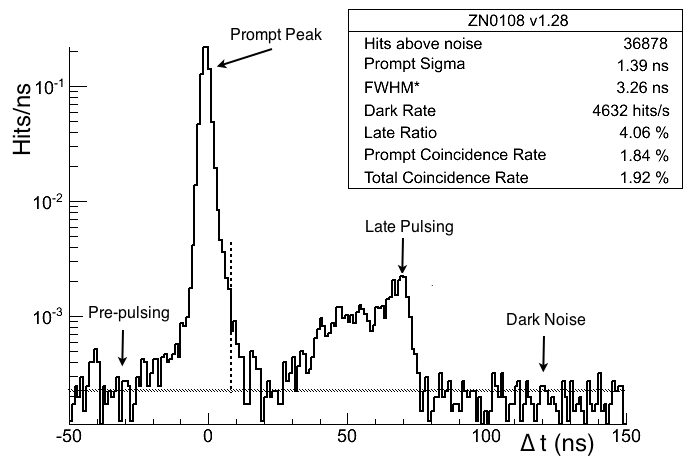}
    \caption{ On the left, all traces with a coincident pulse from one 
	      individual data run are simultaneously plotted.  The
	      red color signifies traces that are within the prompt timing window. The magenta traces signify saturated pulses that were not
	      used in the analysis because it was cut
	      off by the limited dynamic range on the scope. The blue traces signify traces that fall within the late pulsing region, and finally the black
	      traces indicate pulses that are either dark hits or pre-pulses.  The right plot shows a typical $\Delta t$ distribution with the different timing regions labeled.}
    \label{traces_and_time}
\end{figure}

All coincident events taken from a typical data run are shown on the left side
of Figure~\ref{traces_and_time}.  For each event that has a hit in the
Cherenkov source's trigger PMT, we record a full 250~ns waveform for the R11780
being tested.  The timing of this waveform is tied to the timing of the trigger
PMT, which has a FWHM of just 250~ps and thus contributes a negligible
additional jitter to the PMT timing profile. The waveform from the R11780 is
broken into distinct regions and the total number of pulses in each region are 
counted if the charge integral of the pulse is greater than the electronic noise width.  We define the time of a pulse as the time
which the pulse amplitude crosses 20\% of its peak height.  We define a $\Delta t$ for each pulse by subtracting the time of the peak of the pulse from the trigger PMT from the pulse time.

A typical $\Delta t$ distribution is shown on the right in
Figure~\ref{traces_and_time}.  The timing distribution has several clear
features: a nearly-Gaussian prompt peak, a broad feature attributed to late
pulsing that peaks at roughly 75~ns, a continuum of late pulses between the prompt and the late pulsing peaks, a small peak attributed to pre-pulsing at
around 30~ns before the prompt peak, and a uniform distribution of noise hits
throughout the entire distribution.  Late pulsing is caused by elastic
scattering of photoelectrons off the first dynode, where they return later to
the first dynode roughly two cathode-to-dynode transit times later.  The continuum of late pulses is discussed in Section~\ref{sec:high_stats} and pre-pulses
are caused when a photon passes through the glass and directly strikes the first dynode, and we define dark hits to be hits
caused primarily by thermal electron emission from the photocathode and therefore are randomly distributed in time. 

The timing parameters we have measured are:
\begin{itemize}
    \item \textbf{Prompt Sigma:} Width of the prompt timing peak, measured by
fitting a Gaussian in a $\pm$3~ns region around the maximum bin in the $\Delta
t$ distribution.  The position resolution in a water Cherenkov detector depends on this parameter most directly.
    
    \item[] \textbf{FWHM:} The FWHM value of the prompt peak is commonly 
called the `transit time spread' (TTS) and is what PMT manufacturers specify.
Assuming the shape of the prompt peak is described by a Gaussian, the FWHM is taken
to be $2\sigma_{prompt}\sqrt{2ln(2)}$.  
    
    \item[] \textbf{Dark Rate:} Rate of hits that that fall outside the
prompt, the late pulsing, and the pre-pulsing regions.

    \item[] \textbf{Late Ratio:} Fraction of hits in the late pulsing region. 
    Dark hits are subtracted when calculating the ratio.
 The late pulsing region can be PMT-dependent, but we take this region to be
from 5$\sigma_{prompt}$ past the prompt peak to  80~ns after.  Late pulsing is described in more detail in Section \ref{sec:dbl}.
    
    \item[] \textbf{Prompt Coincidence Rate:} Rate of prompt pulses (with
the dark rate subtracted). To ensure the data for a given PMT is in the SPE regime, this value is kept below $5\%$ for all tests.  
    
\end{itemize}

	Tables~\ref{tbl:tresultsstd} and~\ref{tbl:tresultshqe} show the results
for our measurements of these parameters, for standard and high quantum
efficiency versions of the R11780, respectively.  

\begin{table}[htb!]
\renewcommand{\arraystretch}{1.0}
\centering
 \begin{tabular}{c c c c c}
 \hline
  & Mean & Standard Deviation &  Minimum & Maximum\\
 \hline
Transit Time Spread ($\sigma_{prompt}$) (ns)  &  1.37 & 0.15 & 1.20 & 1.6    \\
Late Pulses (fraction)  &  $ 4.48\%$ & $0.32\%$ & $3.93\%$ & $4.92\%$  \\
Noise Rate (Hz) & 3669 & 5110 & 1962 & 16807 \\
Operating Voltage (V) & 1848 & 75 & 1740 & 1920 \\
 \hline
\end{tabular}
\caption{Summary of PMT SPE timing characteristics results for the R11780 standard quantum efficiency PMT. The operating voltage was adjusted to achieve a gain of $1\times10^{7}$, which corresponds to a SPE charge peak of 1.6~pC. All measurements were made at room temperature ($20^{\circ}$ C).}
\label{tbl:tresultsstd}
\end{table}

\begin{table}[htb!]
\renewcommand{\arraystretch}{1.0}
\centering
 \begin{tabular}{c c c c c}
 \hline
  & Mean & Standard Deviation &  Minimum & Maximum\\
 \hline
Transit Time Spread ($\sigma_{prompt}$) (ns)  &  1.29 & 0.14 & 1.16 & 1.52    \\
Late Pulses (fraction)  &  $ 4.3\%$ & $0.35\%$ & $3.6\%$ & $4.8\%$  \\
Noise Rate (Hz) & 4428 & 1897 & 2398 & 8217 \\
Operating Voltage (V) & 1950 & 221 & 1750 & 2500 \\
 \hline
\end{tabular}
\caption{Summary of PMT SPE timing characteristics results for the R11780 high quantum efficiency PMT. The operating voltage was adjusted to achieve a gain of $1\times10^{7}$, which corresponds to a SPE charge peak of 1.6~pC. All measurements were made at room temperature ($20^{\circ}$ C).}
\label{tbl:tresultshqe}
\end{table}

As the tables show, both tubes have outstanding timing resolution, making
precision reconstruction in very large detectors possible.  Their TTS is
substantially better than both the PMTs used in the Super-Kamiokande and SNO
detectors~\cite{sno,superk}.  The higher dark noise rate in the HQE PMTs is not
surprising, given their overall higher quantum efficiency.  

\subsection{Relative Quantum Efficiencies of R11780 Configurations \label{sec:reff}}

	The single photoelectron response of the standard QE and HQE PMTs are similar and are both very good. To estimate the overall gain in quantum efficiency of the HQE configuration compared to the standard version, we tested the relative quantum efficiency of the HQE configuration compared to the standard QE
PMT by running both tube configurations triggering off the same Cherenkov source.  Although the overall particle detection efficiency of a PMT is the product of the quantum efficiency response with its collection efficiency, the ratio of the coincidence rates provides an accurate measure of the relative quantum efficiency since the PMT collection efficiency is the same for both tube configurations.  We compared the relative efficiencies of six HQE tubes to six standard QE tubes. Hamamatsu anticipated about a $52\%$ average increase in quantum efficiency between the HQE and standard QE configurations at 390 nm. 
Using the relative coincidence rates as a measure of the relative quantum efficiencies, we find that the relative quantum efficiency of the HQE version of the
tube is on average $50\%$ higher than the standard quantum efficiency
version as shown in Table~\ref{tbl:eff}.  These results are consistent with Hamamatsu's projections, however, the light which triggered the tested PMTs exhibited a broadband wavelength spectrum. Hamamatsu large-area PMTs are typically sensitive in the approximately 300-600 nm range. The $15\%$ standard deviation in the relative efficiency is dominated by two outlying PMTs.  These outlying PMTs only demonstrated a $32\%$ improvement over the standard tubes. The results of these outlying PMTs could mean either that these particular HQE tubes have a lower peak quantum efficiency than the expected $32\%$, or that the particular standard tubes used for the relative comparison have a higher peak quantum efficiency than the expected value of $21\%$.  Further testing of larger samples of PMTs could settle this.

\begin{table}[htb!]
\renewcommand{\arraystretch}{1.0}
\centering
 \begin{tabular}{c c}
 \hline
& Relative Efficiency\\
 \hline
Tested Pair 1  & $1.56$ \\ 
Tested Pair 2 & $1.49$ \\ 
Tested Pair 3 & $1.66$ \\ 
Tested Pair 4 & $1.64$ \\ 
Tested Pair 5 & $1.32$ \\ 
Tested Pair 6 & $1.32$ \\ 
Average & $1.50$ \\
Standard Deviation &$ 0.15$ \\
 \hline
\end{tabular}
\caption{Observed relative efficiency of the high quantum efficiency configuration of the R11780 PMT compared to the standard quantum efficiency configuration.  The relative efficiencies were measured using using the relative coincidence rates, which is a reasonable metric for a pure single photoelectron source. }
\label{tbl:eff}
\end{table}

\subsection{Absolute Measurement of the R11780 Particle Detection Efficiency \label{sec:eff}}

The particle detection efficiency (PDE) of one 12'' R11780 standard QE PMT was measured. The absolute quantum efficiency is expected to be around $21\%$ at a wavelength of $390$ nm.  The measured particle detection efficiency is less than the absolute quantum efficiency since collection efficiency is not $100\%$ due to photoelectrons missing the first dynode and inelastic back scattering of photoelectrons off the first dynode.  A DC deuterium lamp was focused into a
monochromator. A filter wheel and a 2'' integrating sphere were mounted on the
monochromator output.  The filter wheel contained a neutral density filter, with
reduction factor of roughly 250, a dark frame, and several open positions.  A 1
mm UV/VIS optical fiber was mounted in one port of the integrating sphere,
carrying the light to a dark box.  A monitoring photodiode was mounted in a
second integrating sphere port. The current from the photodiode was measured
using a pico-ammeter.

A NIST calibrated photodiode in the dark box was used to determine the absolute
photon flux emitted by the optical fiber.  The low light level required for
single photon QE measurement is insufficient to drive this photodiode
significantly above its thermal output, so calibration of fiber emission was
done at higher light levels.  The geometry of the integrating sphere ports and
coupling to the fiber were arranged such that with the filter wheel in the open
position the output of the monitoring photodiode was well up in its operating
range, while that for the calibrated photodiode was near its minimum of a few
10s of pA.  The monochromator was scanned in 10 nm steps through wavelengths
between 290 and 500 nm, and the outputs of both photodiodes were recorded in
separate pico-ammeters.

\begin{figure}[hbt!]
 \centering
    \includegraphics[scale=0.65]{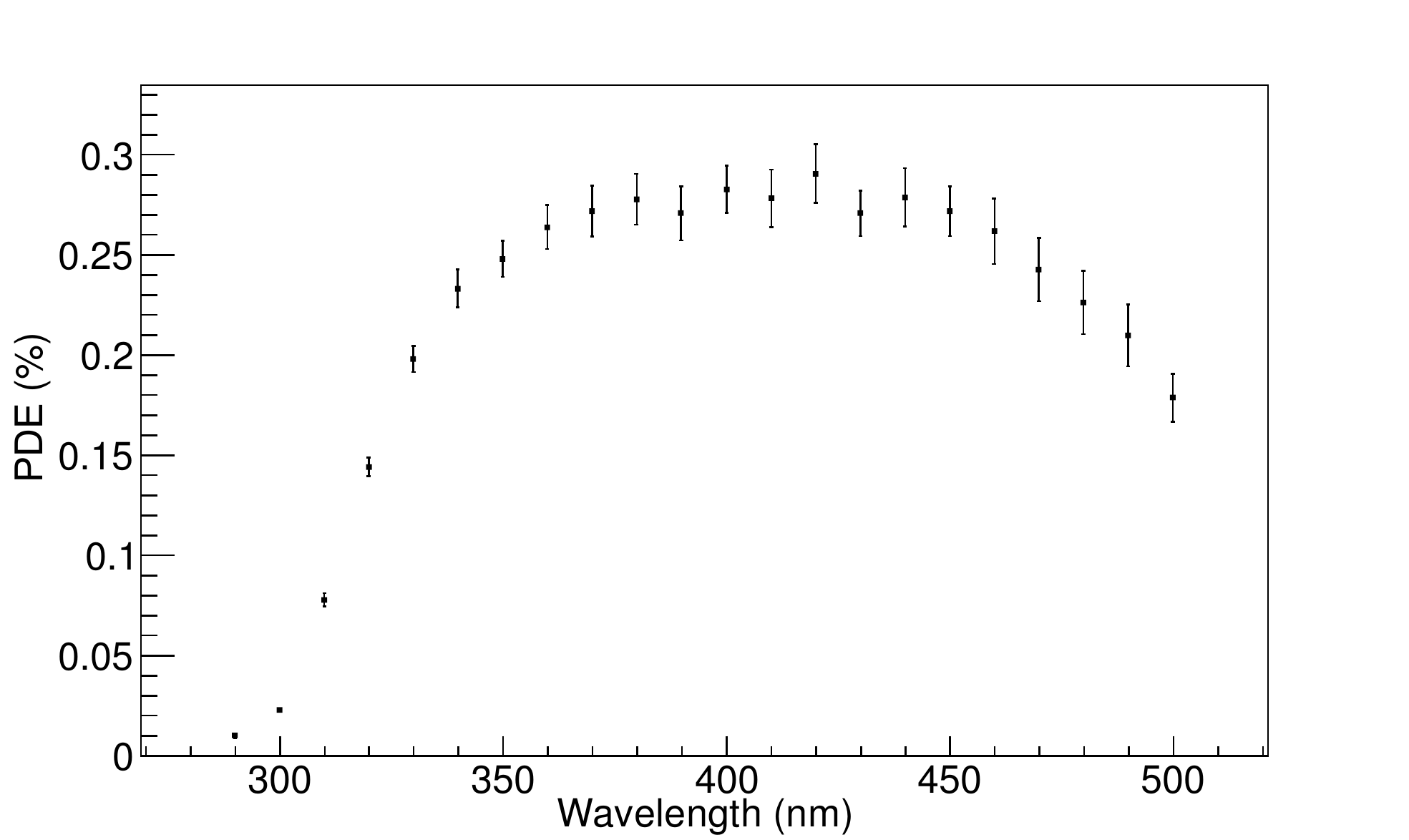}
  \caption{The measured particle detection efficiency of the 12'' R11780 PMT with standard quantum efficiency.}
  	\label{qeplot}
\end{figure}

We then selected the NDF position of the filter wheel, reducing the light by a
factor of 250, and relied on the linearity of the wide dynamic range of the
monitoring photodiode output to provide the reduction factor for calculation of
the fiber emission relative to that measured in the open filter wheel position.  Although the NDF is wavelength dependent, both the integration sphere and the NIST calibrated photodiode are downstream of the NDF and therefore any wavelength dependence does not affect the correlation between the integrating sphere response and the NIST photodiode output.
The equivalent output of the NIST-calibrated photodiode with the NDF in place,
if directly measurable, would be on the fA scale, corresponding to a few kHz of
photons from the fiber end.

The PMT was placed into our dark box with the end of the optical fiber centered
on the photocathode, illuminating a spot of approximately 5 mm$^2$.Ê A voltage
of 1850V was applied to the base, corresponding to the HV used for this tube
during the charge and timing tests described above.Ê The output was connected to
a fast Ortec x10 preamp.  The signal then went to a discriminator with a
threshold set at -400 mV.  The number of pulses below -400 mV were counted for a
period of 10 seconds. We varied the threshold on the discriminator and found
that the background subtracted rate from the PMT was insensitive to
variations. We also looked at the raw traces from the PMT on an oscilloscope and
observed that the real photon hits were much larger than -400 mV. The response
of the PMT was measured from 290 nm to 500 nm in 10 nm steps. At each wavelength
five dark rate and signal rate counts were recorded to insure stability, these
were then averaged and the $signal-dark$ rate for each wavelength was recorded
for each wavelength. The background subtracted rate was typically several kHz.

We then took the background subtracted rates and divided them by the absolute
photon flux for each wavelength, giving the PMT quantum efficiency, shown in the
plot below. The uncertainties are calculated using the one sigma deviation of
all averaged quantities added in quadrature, including the currents from the two
photodiodes and the hit rates measured from the PMT. The errors vary from ~10\%
below 300nm to 3-5\% for the rest of the wavelengths, the higher uncertainty at
290nm is due to a low photon flux at this wavelength.  The uncertainty in the
NIST absolute calibration of the reference photodiode is less than 1\% in this
wavelength range.


We note that the peak PDE of this small spot at the center of the photocathode of
the tested R11780 standard QE PMT is near 28\%, which is significantly higher
than the anticipated PDE averaged over the full photocathode.Ê It is to be expected that the collection efficiency at the center of the photocathode is significantly higher than when averaged over the entire photocathode surface. It is also higher than the $25\%$ average absolute QE of the 10-inch R7081 PMT. Fluctuations in the
relative response over the photocathode have been measured (see Figs. 13 and 14), and this is one possible explanation.Ê Alternatively, the measured PDE can be dependent on the experimental geometry and can include effects of light backscattered from the internal reflective surfaces of
the PMT to the photocathode and the photon-photcathode angle of incidence. The angle of incidence is of particular relevance for PMTs with a curved
face.  In addition, we measure the PDE at a small spot at the center of the photocathode, which differs
from the technique used by the manufacturer and is intended for use in
evaluating an effective PDE in conjunction with full photocathode scans as shown
in Figs.~13 and 14. Finally, this measurement of
the particle detection efficiency may suggest that there are large tube-to-tube variations in the peak quantum efficiency in the initial sample of  R11780 standard QE tubes manufactured by Hamamatsu, with a few tubes including the tested PMT in Figure \ref{qeplot} exhibiting larger than anticipated peak quantum efficiency.  This conclusion is also supported by the data in Table \ref{tbl:eff} since one of the outlying HQE tubes that exhibited a lower than anticipated relative efficiency of $32\%$ over a standard QE tube was in fact compared to the standard QE tube measured in this section. 

\subsection{Late Pulsing and Double Pulsing \label{sec:dbl}}
\label{sec:high_stats}


\begin{figure}[hbt!]
   \centering
    \includegraphics[width=0.75\textwidth]{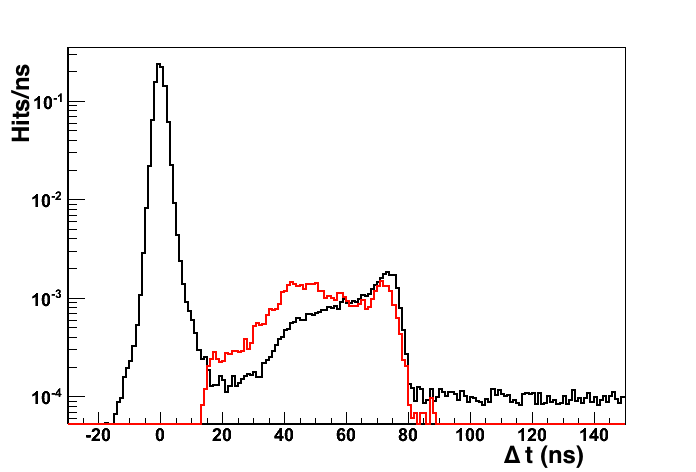}
    \caption{ A high statistics transit time distribution for a 12'' R11780 PMT with high quantum efficiency. The times of the second pulses in double pulse triggers are shown by the dotted red curve.
    }
    \label{figure:doublepulse}
\end{figure}

As was discussed in Section~\ref{sec:TTS}, there is a clear peak near 75~ns in
the transit time spread distribution. This `late pulsing' peak is caused by
elastic scattering of photoelectrons off of the first dynode. After scattering,
the photoelectrons return to the first dynode roughly two cathode-to-dynode
transit times later. There is a related phenomenon known as double pulsing,
which is caused by inelastic scattering off of the first dynode. The inelastic
scatter results in a prompt pulse followed by a second pulse. This second
pulse arrives earlier than is typical for a late pulse because of the energy lost by the
photoelectron due to the scattering.  This second pulse could also be due to `fast afterpulsing',  pulses that arise from light emission from the dynodes themselves, though the time scale of such pulses are typically earlier than the late pulses and double pulses.  Figure \ref{figure:doublepulse} shows
the transit time spread distribution comparing the timing profile of a second
pulse in double-pulse triggered events to the late pulsing distribution for a
R11780 HQE tube. Such characterization of this behavior can be used in algorithms that estimate the
number of detected photoelectrons, or in reconstruction algorithms to ensure
that a double pulse event is not counted as two photons.

\subsection{PMT Position-Dependent Response\label{sec:2dscans}}

The charge and time observables measured in the SPE characterization don't convey any
information about the direction of the incident photon or where it struck the
photocathode.  The optics of the glass and the transmission and absorption of the
semi-conducting photocathode layer can lead to an angular dependence of the
photon detection efficiency. The position-dependent response can in principle be wavelength-dependent, especially for ultraviolet photons which can impart enough transverse momentum to miss the first dynode. The thickness of the PMT glass for the 12-inch tube however, is enough to absorb most UV photons. Optical models of the photocathode layer have been successfully created based on measurements of the complex index of refraction
of the photocathode material. Along with optical models of the
PMT glass, these models provide good predictions for the PMT angular
response~\cite{sno, icecubepmtpaper}.   Such models have been used both in PMT
simulations and in reconstruction algorithms~\cite{dunfordftk}.

\begin{figure}[hbt!]
   \centering
    \includegraphics[width=0.8\textwidth]{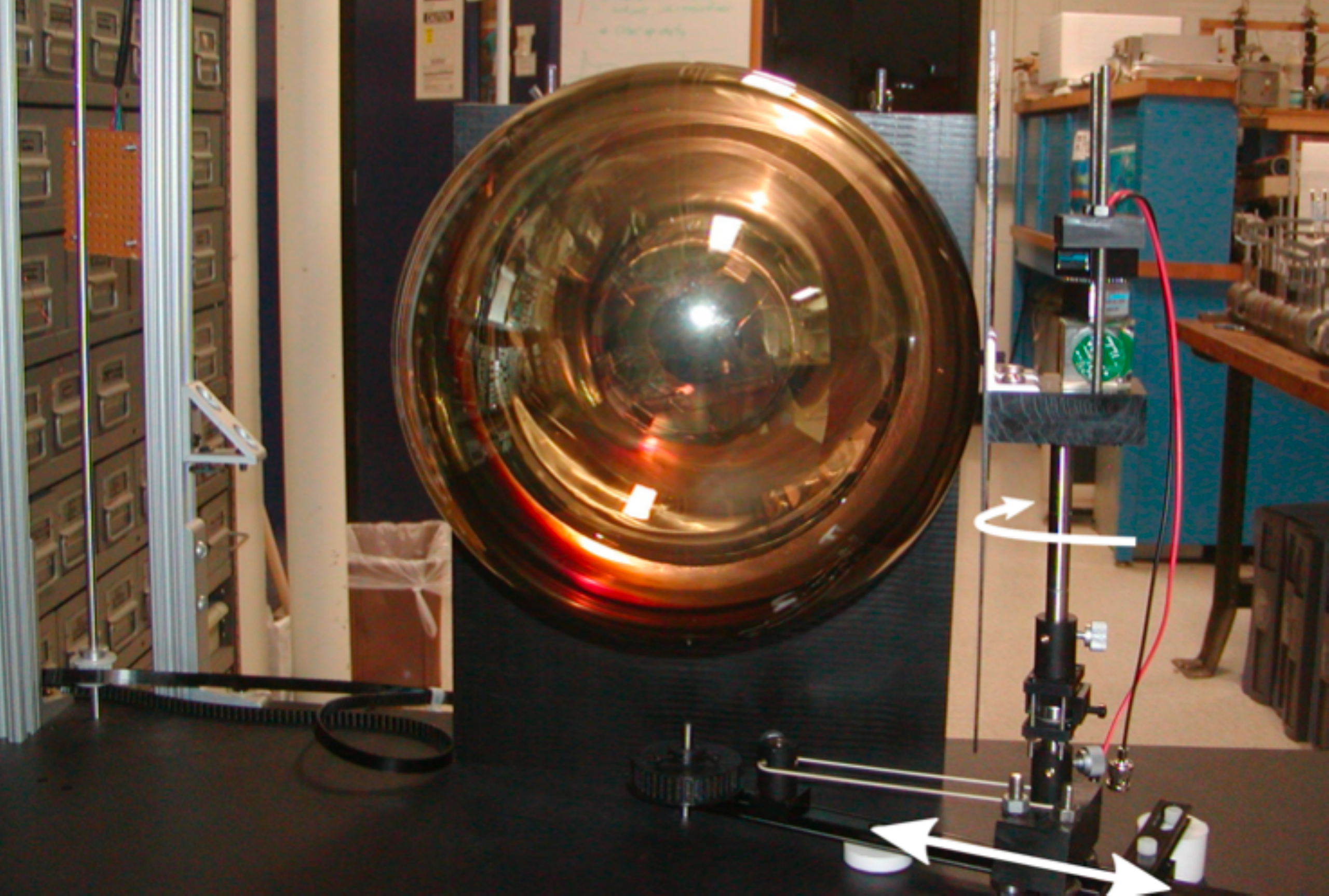}
       \caption{The scanning arm setup used to test the angular dependence of the single photoelectron response of the Hamamatsu R11780 PMT.
    }
    \label{scanningarm}
\end{figure}

\newpage  

\begin{figure}[hbt!]
   \centering
    \includegraphics[width=0.52\textwidth]{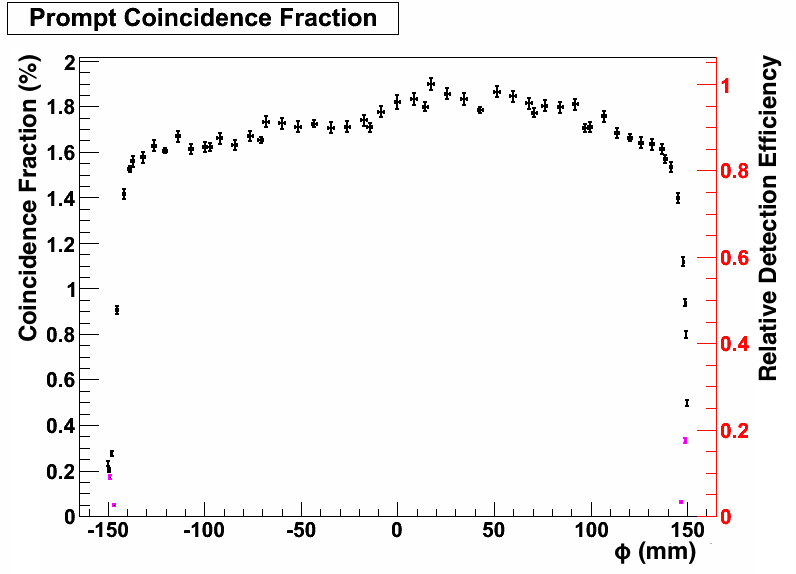}%
    \includegraphics[width=0.54\textwidth]{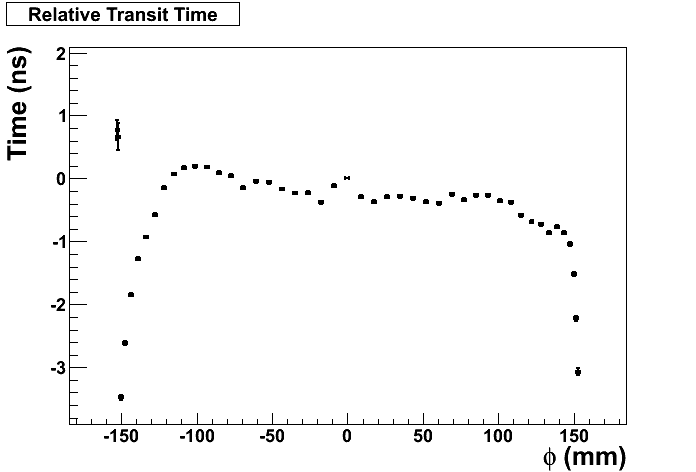}
    \caption{ A one-dimensional scan of the coincidence fraction and relative detection efficiency along the diameter of the PMT is shown on the left,  while position dependent shifts along the diameter in the relative transit time is shown on the right. The scans were made for a Hamamatsu 12-inch R11780 PMT with standard quantum efficiency. }
    \label{1dscan}
\end{figure}

\begin{figure}[hbt!]
   \centering
    \includegraphics[width=0.52\textwidth]{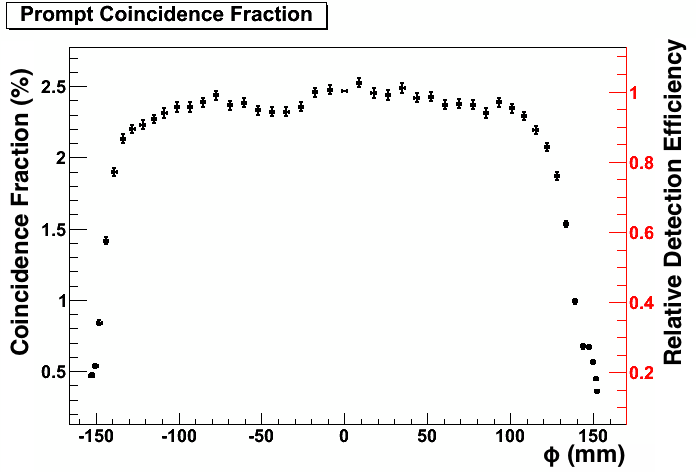}%
    \includegraphics[width=0.52\textwidth]{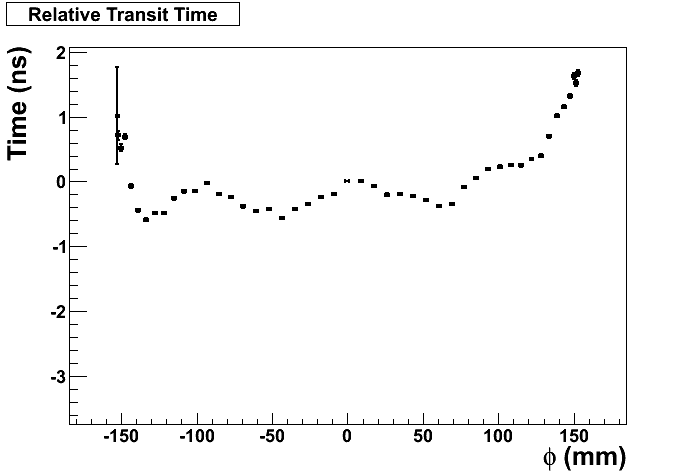}
    \caption{ A one-dimensional scan of the coincidence fraction and relative detection efficiency along the diameter of the Hamamatsu 12-inch R11780 high quantum efficiency PMT is shown on the left,  while position dependent shifts along the diameter in the relative transit time is shown on the right.  The scanned high quantum efficiency PMT has an alternative dynode structure designed to mitigate the large shifts in the mean transit observed in the standard configuration shown in Figure \ref{2dscans}.  }
    \label{1dscanhqe}
\end{figure}

\begin{figure}[hbt!]
   \centering
    \includegraphics[width=0.54\textwidth]{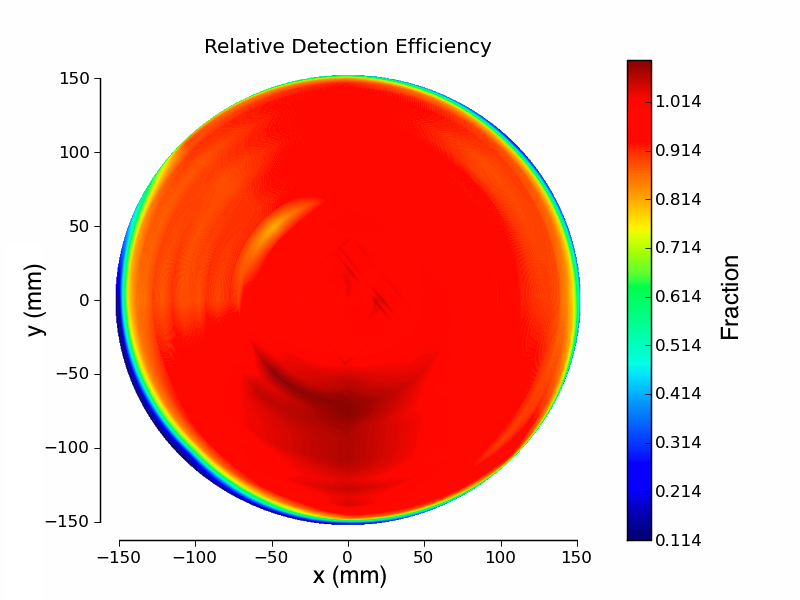}%
    \includegraphics[width=0.54\textwidth]{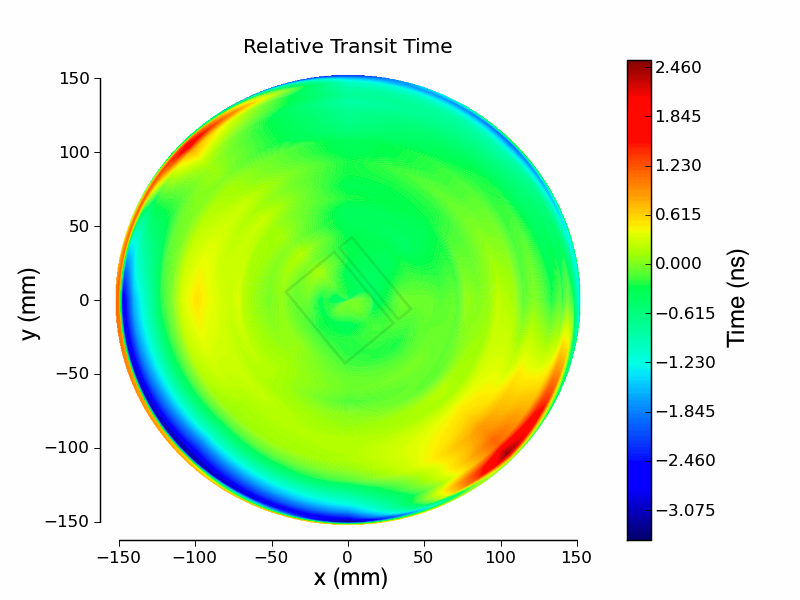}
    \caption{ The position-dependent photon detection efficiency is shown on the left, while position dependent shifts in the median transit time is shown on the right for a Hamamatsu 12-inch R11780 PMT with standard quantum efficiency.  The color indexes are relative to measurements made at the center of the PMT.  
       }
    \label{2dscans}
\end{figure}

\begin{figure}[hbt!]
   \centering
    \includegraphics[width=0.54\textwidth]{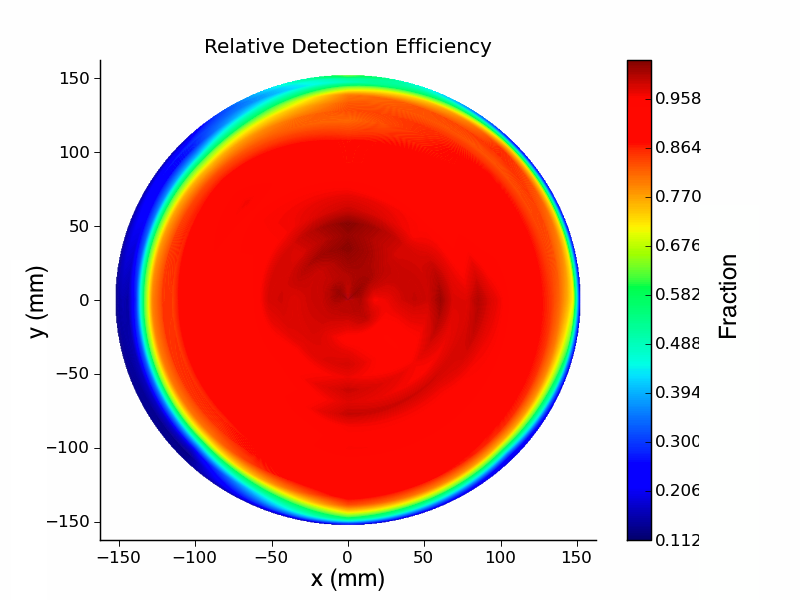}%
    \includegraphics[width=0.54\textwidth]{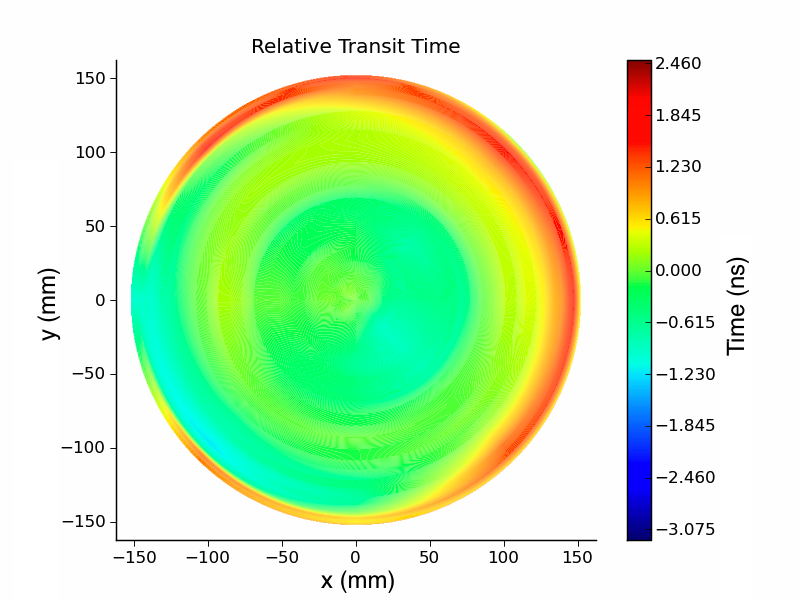}
    \caption{ The position-dependent photon detection efficiency is shown on the left, while position dependent shifts in the median transit time is shown on the right for the high quantum efficiency configuration of the Hamamatsu 12-inch R11780 PMT.  The scanned high quantum efficiency PMT has an alternative dynode structure designed by Hamamatsu to mitigate the large shifts in the mean transit time observed in the standard configuration shown in Figure \ref{2dscans}. The color indexes are relative to measurements made at the center of the PMT.  
       }
    \label{2dscanshqe}
\end{figure}

\begin{figure}[hbt!]
   \centering
    \includegraphics[width=0.6\textwidth]{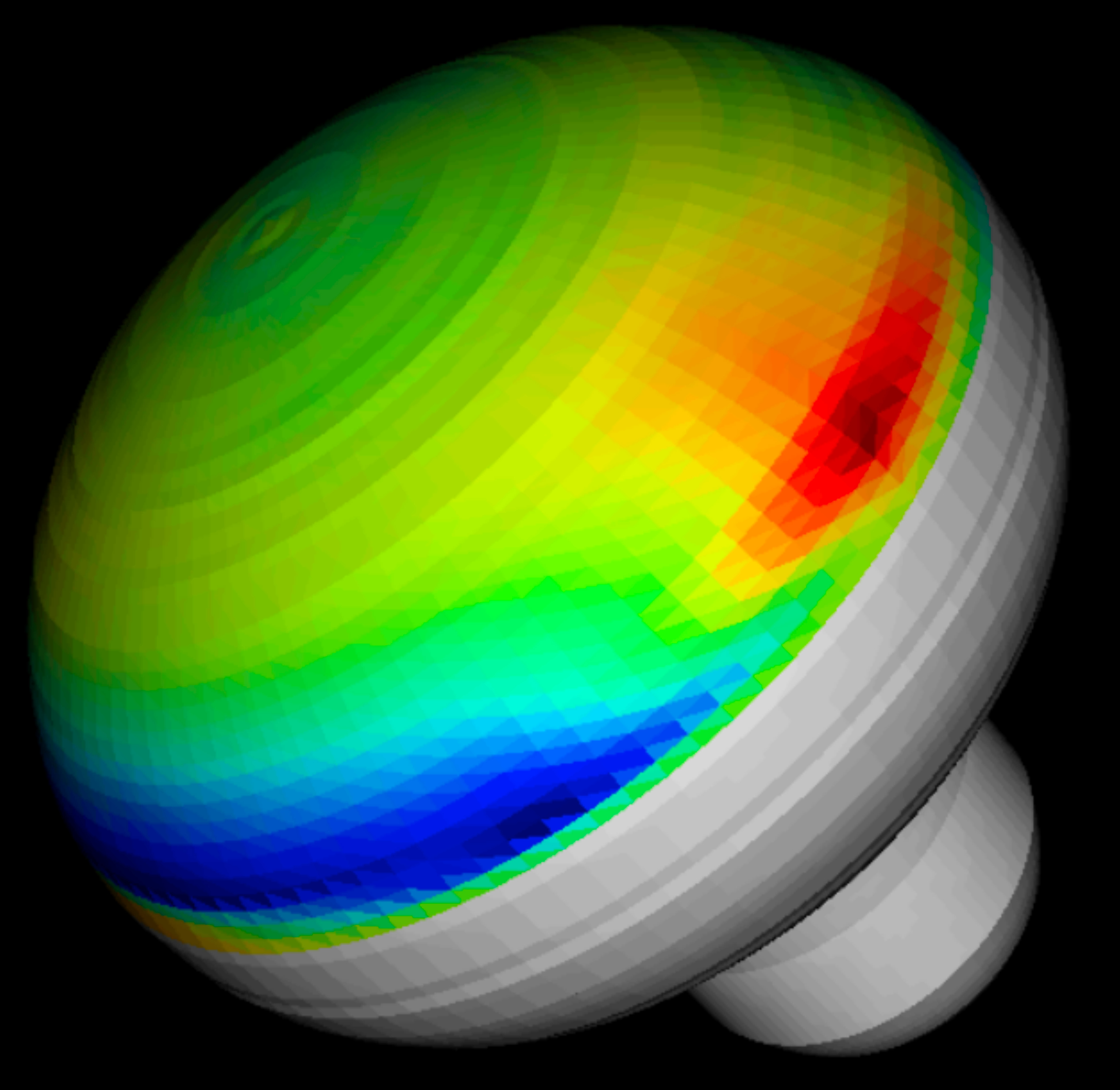}%
        \caption{ The position dependent shifts in the median transit time from Figure \ref{2dscans} is superimposed on a 3d model of the R11780 PMT to demonstrate the size of the regions with large shifts in the median transit time. The color indexes are relative to measurements made at the center of the PMT.  
    }
    \label{fig:3dmap}
\end{figure}

	Beyond the angular dependence, however, there is the possibility of a
position dependence of the photon detection efficiency
or to the overall transit time of a created photoelectron along the photocathode.  Such a dependence can
be due to non-uniformities of the photocathode layer itself or due to variations in the electron optics from the photocathode to
the first dynode. The electron optics can especially affect the position dependence in the observed transit time. The reflective surface within the PMT glass could also affect the position-dependence, however a simulation of the photoelectron response reveals that the impact of the reflective surface on the position-dependence is minor.  A full model of a PMT photon detection efficiency would thus 
include the wavelength-dependence of the quantum efficiency, the
angular-dependence of the efficiency due to the optics of the glass and
photocathode, and any position-dependence of the photon detection efficiency and the transit time.  Although large variations across the
surface of the photocathode can in principle be included in
a complete optical model, such large deviations are not desired and therefore we perform the measurement to characterize the uniformity of the R11780 PMT. 

To measure position-dependence of timing and detection efficiency for the
R11780, we constructed a scanning arm that placed our Cherenkov source at
normal incidence to the PMT face, as shown in Figure~\ref{scanningarm}.  To
ensure we remained in the single photoelectron regime, a pinhole mask was used
to block most of the light except for a pinhole with a diameter of roughly 3 mm.  The position of the source was varied by mounting
the light source, a mask, and the trigger PMT on an optical rail with its pivot
point just below the center of the test PMT.  The entire mounted setup was
controlled by a stepper motor that was operated remotely. The PMT was mounted
at its neck onto a large plastic sheet which itself was mounted onto the
optical rail to ensure proper alignment of source and PMT.  We scanned the
source across the face of the PMT using 50 points across the full diameter,
and then rotated the PMT by 45 degrees between scans. A total of four diameter
scans were performed.

The results of the measurements for a standard quantum efficiency R11780 are 
shown in Figures~\ref{1dscan} and ~\ref{2dscans}.  For the two-dimensional map,
we interpolated along the arc between measurements to produce the contours. In Figure~\ref{fig:3dmap} we show the timing shift map
superimposed on a 3D model of the R11780 to better illustrate the size of the
early-shifted region.

As the figures show, the efficiency of the R11780 is very uniform across
the photocathode, with the exception of the region beyond about 135 mm where it
begins to drop off.  The majority of this region is outside the area that
Hamamatsu specifies for these PMTs.  There is a clear asymmetry in the
lower-efficiency region, which appears to be aligned with the first dynode of
the PMT, indicating that the drop off is more likely caused by electron optics
than a non-uniformity in the photocathode itself.

	We also see in the figures that there are large shifts that are noticeably larger than the transit time spread in the mean transit time as one gets toward the edge of the PMT.  These shifts also appear to be aligned with the dynode structure, and Hamamatsu
has confirmed that their electron optics simulation \cite{anttalk} reproduces these shifts.  Hamamatsu produced a prototype sample of five R11780 high quantum efficiency tubes with an alternative dynode structure designed to mitigate the transit time shifts near the edge of the PMT.  The resulting scans across the PMT surface for one of the new HQE tubes are shown in Figure \ref{1dscanhqe} and Figure \ref{2dscanshqe}.  We note that the large shifts in the relative transit time observed near the edge of the standard quantum efficiency tube are significantly reduced in the new high quantum efficiency configuration with the alternative dynode structure. The cost of the improved timing uniformity is a larger region of decreased photon detection efficiency as shown in the left plot of Figure \ref{2dscanshqe}.  We finally note that the position dependence of the efficiency and the transit time for the new HQE configuration below the region of about 135 mm is extremely similar to the standard quantum efficiency configuration.
  
\subsection{Multiple Photon Contamination\label{sec:sys}}

The results reported in this paper would ideally measure the response of the R11780 PMT to single photons incident to the photocathode.  For an isotropic light source, fluctuations in the number of photons detected by a PMT follow a Poisson distribution and can thus be calculated exactly. For
our source, however, the angular distribution of Cherenkov light and the broad
energy spectrum of the $^{90}$Sr $\beta$ source mean the multi-photon
contamination is much higher. Multiple scattering of the $\beta$ within the
acrylic mitigates this effect somewhat by making the distribution much more
isotropic than a canonical Cherenkov cone, but nevertheless these additional
photons do contaminate our single photoelectron sample.

\begin{figure}[hbt!]
   \centering
     \includegraphics[width=0.485\textwidth]{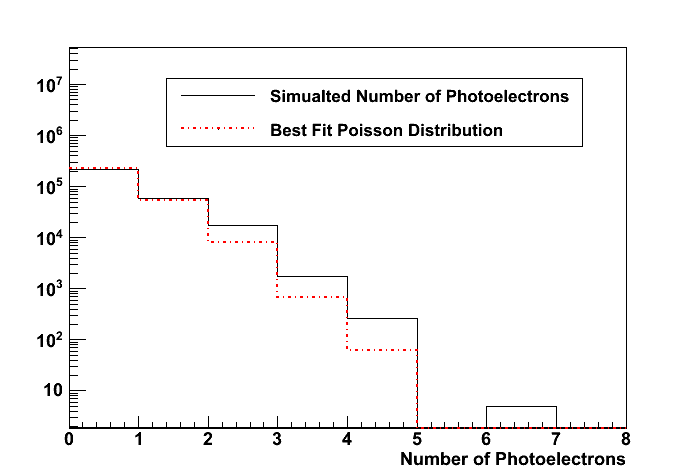}	
       \includegraphics[width=0.485\textwidth]{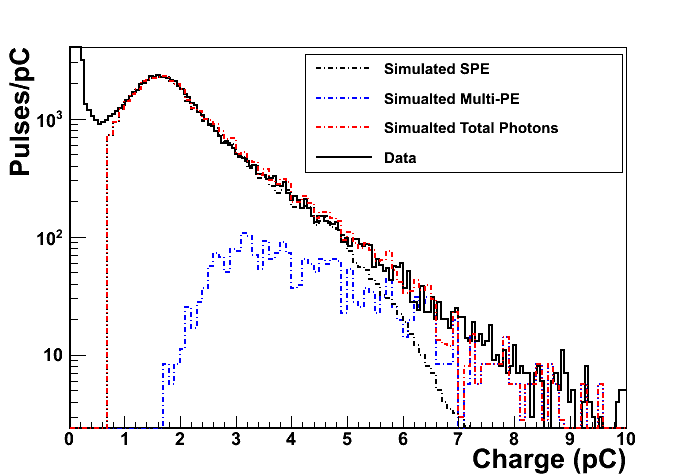}
        \caption{ The left plot shows the distribution of detected number of photoelectrons as predicted by simulation. A Poisson distribution was fit to the photoelectron distribution and is also shown. The simulation was performed using the Reactor Analysis Tool (RAT) software. The right plot shows the measured single photoelectron charge spectrum for a R11780 high quantum efficiency PMT compared with the simulated single photoelectron charge spectrum along with the predicted multiple photon contamination. The simulation predicts that multiple photons contribute $6.5\%$ to the total number of pulses, in comparison to a $3\%$ contribution from a Poisson distribution. The Monte Carlo simulation does not include the effects of afterpulsing, since after pulsing is PMT specific and can not be accurately predicted.
    }
    \label{fig:mpe}
\end{figure}

\begin{table}[htb!]
\renewcommand{\arraystretch}{1.0}
\centering
 \begin{tabular}{c c c c}
 \hline
Detected photons & Simulated $\%$ of total  & Poisson $\%$ of total & Simulation / Poisson \\
\hline
$2$ or more & 6.5$\%$ & $3\%$ & 2.2 \\ 
 2 & 5.8$\%$ & $2.7\%$ & 2.15 \\ 
 3 & 0.67$\%$ & $0.23\%$ & 2.5 \\ 
 4 & 0.087$\%$ & $0.021\%$ & 4.1 \\ 
 \hline
\end{tabular}
\caption{The multiple-photon contamination contribution from the simulation and a Poisson distribution is tabulated for various numbers of detected photons.  The simulation was performed using the RAT software. The Poisson distribution was taken from a fit to the simulated photoelectron distribution in the left plot of Figure \ref{fig:mpe}. }
\label{tbl:mpe}
\end{table}

To estimate the level of multi-photon contamination, we developed a Monte Carlo simulation of the SPE characterization experimental setup using the RAT \cite{rat} framework, which is used for the simulation of the SNO+ and MiniCLEAN experiments.  The Monte Carlo simulation incorporated a full description of the geometry, optics, timing, and quantum efficiency response of the R11780 PMT.  The single photoelectron response for the R11780 implemented in the simulation was taken from the measured data.  A high statistics Monte Carlo simulation of two million events was generated in order to estimate the contribution of  two or more incident photons on the simulated charge spectrum for the R11780 12-inch PMT.  The simulated number of detected photoelectrons is compared to a fit of a Poisson distribution to the photoelectron spectrum on the left plot of Figure \ref{fig:mpe}.  The simulated multi-photon contribution from two or more detected photoelectrons is $6.5\%$ of the total number of photoelectrons, which is higher compared to the Poisson contribution of $3\%$.  The detected multiple-photon contamination from simulation and a Poisson distribution is tabulated in Table \ref{tbl:mpe} for different multiplicities of detected photons.  We note that the deviation of the simulation from the Poisson prediction increases for higher multiplicities of detected photoelectrons.  

The resulting simulated charge spectrum is compared to data on the right plot of Figure \ref{fig:mpe}.  The simulation predicted that $6.5\%$ of all simulated pulses arose from multiple-photon events, which is compatible with the measured high-charge tail of the SPE charge distribution.  We  note the possibility of bias in the right plot of Figure \ref{fig:mpe} since the simulated single photoelectron response is taken from the data, which is known to have contamination from multiple photoelectrons.  However, the simulation shows that the charge distribution arising from the contribution of multiple photons is well away from the single photoelectron peak and shows up predominately in the high charge tail.

Our simulation has shown that the chosen SPE characterization parameters,
with the exception of the high charge tail, are insensitive to the multi
photon contamination. So although our Chrenkov source is not a pure
Poisson source, the amount of contamination is acceptable for the SPE
characterization and our parameters are indicative of the true SPE
response.


\section{Conclusions}
\label{conclusions}

The 12'' R11780 PMT from Hamamatsu is an excellent candidate PMT for very
large-scale water Cherenkov or scintillator detectors.  The SPE charge and 
timing response are excellent, and the high quantum efficiency version is a 
cost-effective solution that increases the detected photon yield on average by about 50\%.  A
two-dimensional scan of the standard quantum efficiency version of the R11780 PMT revealed that the charge and timing
response is uniform across most of the photocathode surface, although there are noticeable timing shifts near the
edge of the PMT. Further research and development work by Hamamatsu resulted in an alternative dynode structure designed to resolve the problem. A two-dimensional scan of the high quantum efficiency version of the R11780 PMT with the alternative dynode structure revealed that the observed timing shifts have been mitigated. Finally, a simulation of the single photoelectron experimental setup shows that contamination from multiple photons is relatively small and therefore the results reported here are indicative of the true SPE response of the Hamamatsu R11780 PMT. 

\section*{Acknowledgments}

We acknowledge the support from the following agencies:\\
U.S. National Science Foundation-Physics Division,\\
U.S. Department of Energy

\bibliographystyle{elsarticle-num}
\bibliography{PMTTestingPaper_Submit}
\end{document}